\documentclass[twocolumn,nopacs,amsmath,amssymb,10pt]{revtex4}

\usepackage{graphicx}
\usepackage{bm}
\usepackage{color}
\usepackage{subfigure}

\begin{document}
\title{Matter waves in atomic-molecular condensates with Feshbach resonance management}
\author{F. Kh. Abdullaev$^{1,2}$}
\author{M. \"{O}gren$^{3,4}$}
\author{J. S. Yuldashev$^{1,2}$}
\affiliation{$^1$ Physical-Technical Institute, Uzbek Academy of Sciences, 100084, Tashkent, Uzbekistan.}
\affiliation{$^2$ Theoretical Physics Department, National University of Uzbekistan, Tashkent, Uzbekistan.}
\affiliation{$^3$ School of Science and Technology, \"{O}rebro University, 70182 \"{O}rebro, Sweden,} 
\affiliation{$^4$ Hellenic Mediterranean University, P.O. Box 1939, GR-71004, Heraklion, Greece.}

\date{\today}

\begin{abstract} 
The dynamics of matter waves in the atomic to molecular condensate transition with a time modulated atomic scattering length is investigated. 
Both the cases of rapid and slow modulations are studied. 
In the case of rapid modulations, the average over oscillations for the system is derived.
The corresponding conditions for dynamical suppression of the association of atoms into the molecular field, or of second-harmonic generation in nonlinear optical systems, are obtained. 
For the case of slow modulations, we find resonant enhancement in the molecular field. 
We then illustrate chaos in the atomic-molecular BEC system. 
We suggest a sequential application of the two types of modulations, slow and rapid, when producing molecules.
\end{abstract}

\maketitle

\section{Introduction}
The dynamics of nonlinear waves in quadratic nonlinear media with periodic modulations in time of the parameters is an active area for investigations~\cite{Buryak,Malomed}. 
The interest is connected with the possibility of quasi-phase-matching in such media and the ability to control the interactions of the waves. 
The main type of modulation that attracts interest, is the case of periodically poling crystals with rapid modulations of the quadratic nonlinearity parameter along the direction of propagation. 
The existence of quasi-phase-matching~(QPM) for competing nonlinearities in solitons has been shown~\cite{Clausen} and new types of all-optical switching~\cite{Kobyakov} have been demonstrated. 
Another interesting case is the periodic modulation of mismatch parameters, which can be realized in a nonlinear optical medium~\cite{Trillo}, or for matter waves in atomic-molecular condensates~\cite{Vardi}.

Recently soliton dynamics in quadratic nonlinear optical media with a Kerr nonlinearity varying along the direction of propagation have been investigated~\cite{opt_spectr}.   
It was shown by numerical simulations that the resonant response of the amplitude oscillations of the solitons appears at the frequency of modulations for the Kerr nonlinearity and is equal to the frequency of modulation for the uniform phase of the solitonic solution.  
A second system described by this type of model is an atomic-molecular condensate~\cite{Drummond,VKD,Timmermans1,Yurovsky_1999_2000,Cusak,OgrenPRA2010}, while varying in time the atomic scattering length. 
The latter case is implemented by the so-called Feshbach resonance management~\cite{Abd,Yurovsky_1999_2000,Malomed,AKS}, and it can be realized for example by a variation of an external magnetic field near the resonance value~\cite{Petrov}.

In this work, we will investigate analytically and numerically the propagation of continuous waves in a quadratic nonlinear ($\chi^{(2)}$) system with an additional nonuniform cubic (Kerr) nonlinearity. 

The structure of the article is the following: 
In section~\ref{sec_Model} we describe the model for propagation of the fundamental and second harmonics in a medium with competing cubic and quadratic nonlinearities,
where the Kerr nonlinearity can be periodically varying along the direction of evolution. 
We start with a description of the unperturbed $\chi^{(2)}$ system with a Kerr nonlinearity and its solution for physically valid parameters in section~\ref{sec:unperturbed}.  
In section~\ref{sec_CW} we then investigate the evolution of continuous waves in the system. 
We derive the averaged over rapid modulations of the cubic nonlinearity in the direction of evolution. 
In section~\ref{sec_Strong} we study in detail the corresponding management and
the efficiency in the generation of the second harmonic (the molecular field) is investigated. 
We further investigate resonances in a $\chi^{(2)}$ system with periodic Kerr nonlinearity in the case of slow modulations in section~\ref{sec_slow_modulation}.
Finally in section~\ref{sec_Conclusion} we conclude the study.

\section{The model} \label{sec_Model}

The system, describing the propagation of the fundamental harmonics (FH) and second harmonics (SH) in a quadratic nonlinear medium with a cubic
nonlinearity, has in standard optics dimensionless variables the form~\cite{Buryak} 
\begin{eqnarray}\label{eq1}
iu_z+ u_{xx} + \gamma(z)\lvert u\rvert^2u+u^*v&=&0,  \nonumber \\
iv_z+ \frac{1}{2}v_{xx} + qv+\frac{u^2}{2}&=&0.
\end{eqnarray}
Here $u,v$ are the fields of the FH and the SH respectively. 
For an atomic-molecular BEC system they are instead the atomic and molecular fields~\cite{Drummond}. 
In optics Kerr nonlinearities may also be included in the second harmonics and for the cross terms~\cite{OleBang}. 
Below we will consider the model~(\ref{eq1}), keeping in mind mainly the atomic-molecular BEC system.

The related system of coupled Gross-Pitaevskii like equations with conversion terms describing the atomic-molecular BEC is in physical units
\begin{eqnarray}\label{gpe}
i\hbar\psi_{a,T}&=&-\frac{\hbar^2}{2m_a}\psi_{a,XX}+ \sum_{j=a,m} g_{aj}|\psi_j|^2\psi_a  - \hbar G_{am}\psi_a^{\ast}\psi_m ,\nonumber \\
i\hbar\psi_{m,T}&=&-\frac{\hbar^2}{2m_m}\psi_{m,XX}+ \hbar \delta\omega\psi_m \nonumber \\
&+& \sum_{j=a,m} g_{jm}|\psi_j|^2\psi_m - \hbar G_{am}\frac{\psi_a^2}{2},
\end{eqnarray}
where $m_m=2m_a$ are the masses, $\hbar \delta\omega $ is the energy detuning, $g_{aa}$, $g_{mm}$, $g_{am}$, and $G_{am}$ are one-dimensional parameters for the atom-atom, molecule-molecule, and the atom-molecule interactions~\cite{OgrenPRA2010}, with (e.g.) $g_{aa}=2\hbar \omega_\perp a_s$, where (e.g.) $a_s$ is the atomic scattering length, and $\omega_\perp$ is a strong enough transverse trapping frequency for the system to be quasi-one-dimensional~\cite{Yurovsky2008}.
Finally the parameter $G_{am}$ is the strength of the atom-molecule conversion, while effects of elastic collisions involving molecules, i.e., $g_{mm}$ and $g_{am}$ in Eq.~(\ref{gpe}), will be neglected here~\cite{OgrenPRA2010}.

The dimensionless form, i.e., Eqs.~(\ref{eq1}), is obtained by the following change of variables in the system~(\ref{gpe}):
\begin{eqnarray}
z=T\omega_{\perp}, x=\frac{ \sqrt{2}X }{l_a}, l_a=\sqrt{\frac{\hbar}{m_a\omega_{\perp}}}, \nonumber \\
u=\frac{G_{am}}{\omega_{\perp}} \psi_a,
v=\frac{G_{am}}{\omega_{\perp}}\psi_m,\gamma =-\frac{\omega_{\perp} g_{aa}}{\hbar G_{am}^2}, q=-\frac{\delta\omega}{\omega_{\perp}} \nonumber.
\end{eqnarray}
Time dependent variations of the atomic scattering length $a_s$ and therefore the parameter $\gamma$, can be obtained as mentioned before, by the variations in time of an external magnetic field near a resonant value, the so-called Feshbach resonance management technique~\cite{KaganPRA1996, KaganPRA2007, SaitoPRL2003, AbdullaevPRA2013,Inouye}.
Solitons in an atomic-molecular BEC system with adiabatically tunable interactions were investigated recently in~\cite{Wang}.

In this work, we will investigate rapid and slow periodic variations of the cubic nonlinearity. 
The modulations are taken in the form
\begin{equation}\label{eq3}
\gamma= \gamma_0 + \gamma_1\cos(\omega z),
\end{equation}
where $\omega$ is a dimensionless driving frequency.
The strong management case corresponds to $\gamma_1 \sim \omega \sim 1/\epsilon, \epsilon \ll 1.$ 
The weak management case corresponds to $\gamma_1 \sim O(1),\omega \sim 1/\epsilon$.
To study the dynamics in the case of rapid modulations, we will derive the averaged equations. For slow modulations we will pay special attention to the case of resonant modulations, i.e., when the frequency of the modulation $\omega$ is equal to the frequency for the transformation of the FH into the SH and vice versa.

\section{Description of unperturbed $\chi^{(2)}$ system with Kerr nonlinearity}\label{sec:unperturbed}

In this section we consider the unperturbed system of Eq.~(\ref{eq1}) describing propagation of continuous waves (CW) ($u_{xx} = v_{xx}=0$), i.e., with $\gamma_0 \neq 0$ and $\gamma_1 = 0$. 
The dynamical behavior of such a system without cubic nonlinearity was studied by Bang \textit{et al.}, who explored stationary solutions and self-trapping in a discrete quadratic nonlinear system~\cite{Bang}, and also with quasi-phase-matching induced nonlinearity, investigated in~\cite{Kobyakov}. 
In atomic-molecular BEC systems, the properties of homogeneous solutions in the case when atom-atom, and atom-molecule interactions are neglected were considered in~\cite{Timmermans}. 

We apply conventional normalizations for the amplitudes, the direction of amplitudes, and the mismatch, as 
\begin{equation}\label{eq:normalizations}
u=\sqrt{I}\rho e^{i\phi}, \,\, v=\sqrt{I}\mu e^{i\psi}, \,\, Z=z/L, \,\, \kappa=qL/2, 
\end{equation}
where  $I=\lvert u\rvert^2+2\lvert v\rvert^2$ is the conserved total intensity (i.e., $\rho^2+2\mu^2=1$). 
For the atomic-molecular BEC system, it has the interpretation of the total number of particles.

After inserting the normalizations into Eq.~(\ref{eq1}), the system of differential equations takes the following form in the new variables:
\begin{eqnarray}\label{2:eq2}
i\rho_Z-\rho \phi_Z+L\sqrt{I}\rho\mu e^{-i(2\phi-\psi)} + LI\gamma_0 \rho^3= 0, \nonumber \\
i\mu_Z-\mu \psi_Z+2\kappa \mu + L\frac{\sqrt{I}}{2}\rho^2 e^{i(2\phi-\psi)} = 0.
\end{eqnarray}   
By defining the parameters $\beta = L\sqrt{I}$, $\Upsilon = LI\gamma_0$, and $\theta=2\phi-\psi$, and separating the real and imaginary parts, we get the following expressions for the amplitudes and phase:
\begin{equation}\label{2:eq3}
\rho_Z=\beta \rho \mu \sin\theta,  \,\,\,\,\,\,\,\, \mu_Z=-\frac{\beta}{2}\rho^2 \sin\theta, 
\end{equation}
and
\begin{equation}\label{2:eq4}
\theta_Z=2\phi_Z-\psi_Z=-2\kappa+ 2\Upsilon \rho^2 + \left( 2\beta\mu-\frac{\beta\rho^2}{2\mu} \right) \cos\theta.
\end{equation}
By defining the parameters $\lambda = \Upsilon/\beta$, $\sigma=\kappa/\beta$, and $w = \mu^2$, we can derive a Hamiltonian system with two conjugate degrees of freedom,
\begin{equation}\label{2:eq5}
\dot{w} = -(1-2w)\sqrt{w}\sin(\theta)=\frac{\partial H}{\partial \theta},
\end{equation}
\begin{equation}\label{2:eq6}
\dot{\theta} = -2\sigma + 2\lambda(1-2w)-\left( \frac{1-6w}{2\sqrt{w}}\right)\cos(\theta)=-\frac{\partial H}{\partial w},
\end{equation}
where an overdot denotes differentiation with respect to $\xi = \beta Z$, and $H$ is given by 
\begin{equation}\label{2:eq7}
H = 2\sigma w - 2\lambda(w-w^2)+(1-2w)\sqrt{w}\cos(\theta).
\end{equation}

Applying the identity $\sin^2(\theta)+\cos^2(\theta)=1$ to Eqs.~(\ref{2:eq5}) and~(\ref{2:eq6}), these equations can be written in the form of a single differential equation for the relative intensity, $w$, of the second harmonic,
\begin{equation}\label{2:eq8}
\dot{w}^2 + P(w) = 0,
\end{equation}
which is equivalent to the dynamical equation describing a classical particle moving in a potential $P(w)$. 
The potential $P(w)$ is in the form of a quartic polynomial,
\begin{eqnarray}\label{2:eq9}
P(w) = D^2w^4 - 4(1+DC)w^3 \nonumber \\
+4(1-\frac{1}{2}HD+C^2)w^2 -(1-4HC)w + H^2,
\end{eqnarray}
where $C = \lambda -\sigma$ and $D = 2\lambda$.

\begin{figure}[ht!]
\begin{center}
\includegraphics[scale=0.5]{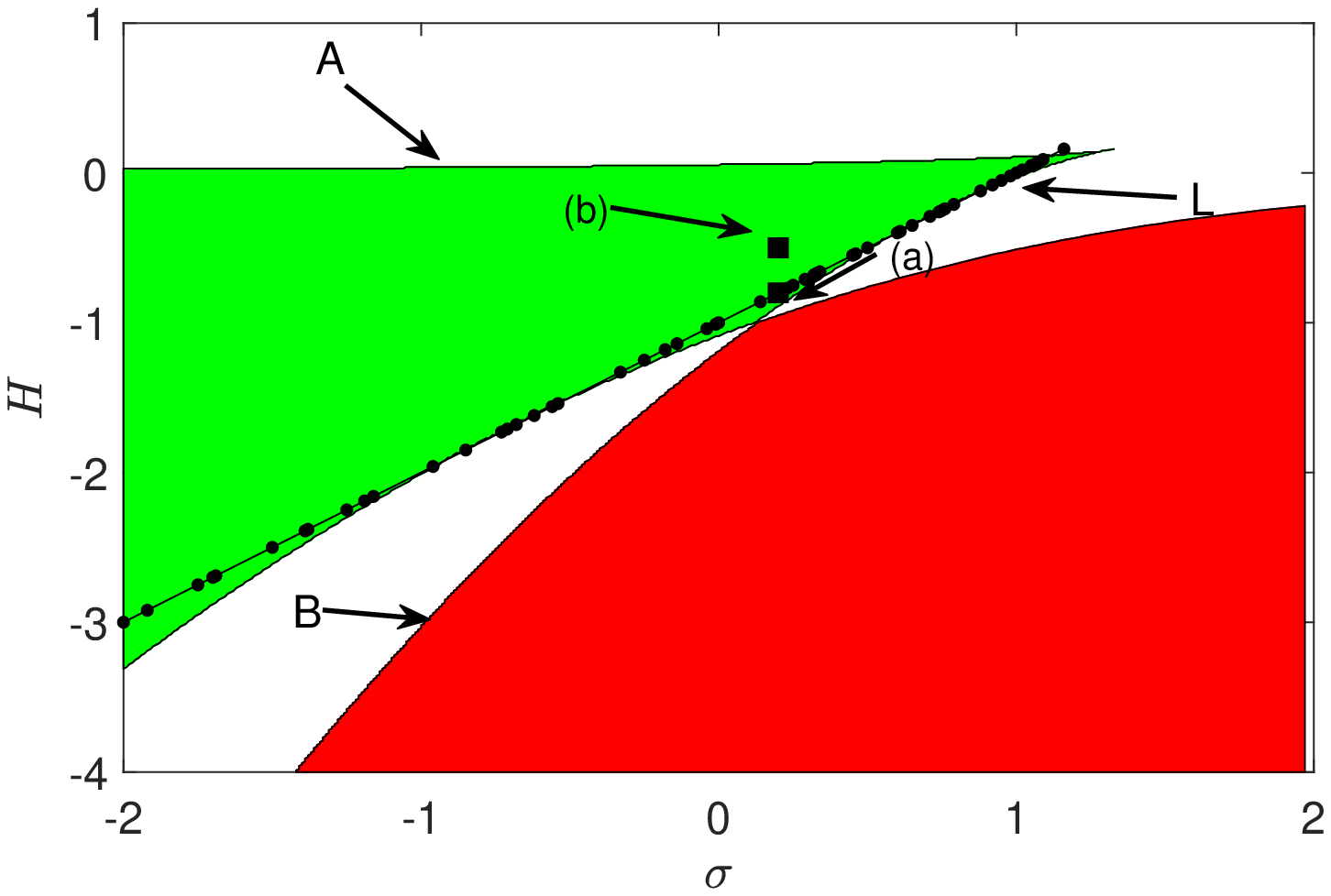}
\includegraphics[scale=0.5]{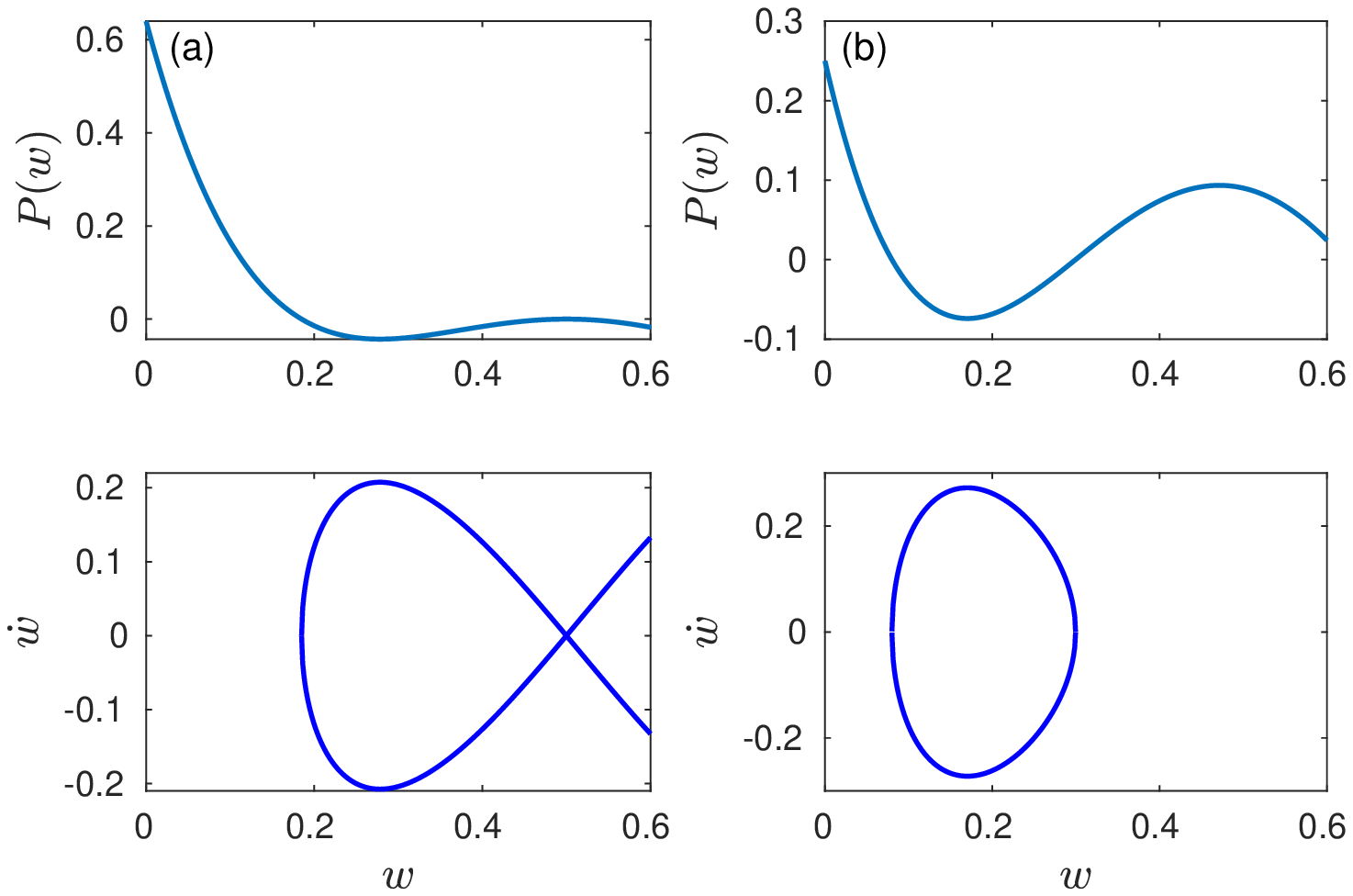}
\caption{(Color online) 
(Upper) Regions of the $(\sigma, H)$ plane in which the quartic equation $P(w)=0$ given by Eq.~(\ref{2:eq9}) has four real roots, region A (green); two real roots (white region); and no real root region B (red). 
Large (black) dots on the line ($L$) indicates separatrices in the phase space, when two of the four roots are equal. 
(Lower) The potential and corresponding phase space for the two (black) squares at different regions in the $(\sigma, H)$ plane (see the upper part). The values are $\sigma=0.2$, $H=-0.8$ for (a), and $\sigma=0.2$, $H=-0.5$ for (b).}
\label{fig:2.1}
\end{center}
\end{figure}

In Fig.~\ref{fig:2.1} (upper), a classification of the roots of the quartic equation $P(w)$ in the ($\sigma, H$) plane with $\lambda=2$ is illustrated. 
In the upper left shaded (green) region (A), $P=0$ has four real positive roots. 
In the bottom right shaded (red) region (B) there are four complex roots.
Finally, there are two complex, and two real roots, in the white regions. 
Large (black) points on the line ($L$) corresponds to the case with four real roots, of which two are equal.
Such a (black) point, corresponds to a separatrix in the phase space, which describes a complete transformation from the power of the FH into the SH [case (a)]. 
The potential $P(w)$ and its corresponding phase portraits are shown in Fig.~\ref{fig:2.1} (lower) for $\sigma = 0.2$ and two different values of $H$, corresponding to each regions in the $(\sigma, H)$ plane. 
In general, Eq.~(\ref{2:eq8}) has solutions in the physically valid regions, the upper left (A) region, and at the large (black) dots on the line ($L$). 
In the physical region (A) of the $(\sigma, H)$ plane, Eq.~(\ref{2:eq8}) has a solution $w(\xi)$ which is a periodic function, determined by the four real roots $w_0 < w_1 < w_2 < w_3$, which oscillates between the two lowest roots $w_0$ and $w_1$. 
The solution can be explicitly obtained by integrating Eq.~(\ref{2:eq8}) according to~\cite{ByrdAndFriedman}

\begin{eqnarray}
\xi = \int_0^{\xi} d\Xi = \int_0^{w(\xi)}\frac{dW}{\pm\sqrt{-P(W)}} \nonumber \\ 
= \frac{1}{\pm D}\int_0^{w(\xi)}\frac{dW}{\sqrt{(W-w_3)(W-w_2)(W-w_1)(W-w_0)}} \nonumber \\ 
= \pm \frac{1}{ND}\textnormal{sn}^{-1}\Bigg[\sqrt{\frac{(w_3-w_1)(w-w_0)}{(w_1-w_0)(w_3-w)}}|k \Bigg], \label{eq:sn_inverse}
\end{eqnarray}
where $\textnormal{sn}(.|.)$ is a Jacobi elliptic function.
Hence, we have the following solution from Eq.~(\ref{eq:sn_inverse}):
\begin{equation}\label{eqfor_w}
w(\xi) = \frac{w_3 a \textnormal{sn} ^2(r\xi |k) + w_0}{a \textnormal{sn} ^2(r\xi |k) + 1},
\end{equation}
where 
$$
r = ND = \lambda\sqrt{(w_3-w_1)(w_2-w_0)},  
$$
and
$$
k = \sqrt{\frac{(w_3-w_2)(w_1-w_0)}{(w_3-w_1)(w_2-w_0)}}, \: a = \frac{w_1-w_0}{w_3-w_1}.
$$

\section{Averaged equations for strong management} \label{sec_CW}

In  the case of rapid modulations for a strong management case,
we consider modulations of the cubic nonlinearity of the form  
$$\gamma = \gamma_0 + \gamma_1 f(\zeta) = \gamma_0 + \frac{1}{\epsilon} f \left( \frac{z}{\epsilon} \right),$$ 
where 
$f$ is a periodic function of $\zeta =z/\epsilon$, and $\epsilon \sim 1/ \omega \ll 1$.
Our goal is to derive the corresponding average over the rapid modulation of the system~(\ref{eq1}). 
We will use the following transformation to a new field for the FH:
\begin{equation}\label{eq2}
u=\bar{u}e^{i\varGamma(z)\lvert \bar{u}\rvert^2}, \: v = \bar{v},
\end{equation}
where $\Gamma(z)$ is the anti-derivative of $\gamma_1f(z)$, i.e., $\Gamma_z=\gamma_1 \cos(\omega z)$; see Eq.~(\ref{eq3}).
This transformation allow us to exclude strong rapidly varying terms from the mean-field equations~\cite{Zharnitsky,AAG}.
We study here the propagation of continuous waves, i.e., when $u_{xx}=v_{xx}=0$, in a media with a periodically varying Kerr nonlinearity. 
Substituting Eq.~(\ref{eq2}) into Eqs.~(\ref{eq1}), we obtain
\begin{eqnarray}
i\bar{u}_ze^{i\varGamma(z)\lvert \bar{u}\rvert^2}-\varGamma(z)\lvert \bar{u}\rvert^2_z \bar{u} e^{i\varGamma(z)\lvert \bar{u}\rvert^2}+\bar{u}^*\bar{v}e^{-i\varGamma(z)\lvert \bar{u}\rvert^2}  \nonumber \\
+\gamma_0\lvert \bar{u}\rvert^2 \bar{u}e^{i\varGamma(z)\lvert \bar{u}\rvert^2}=0, \nonumber \\
i\bar{v}_z+q\bar{v}+\frac{\bar{u}^2}{2}e^{i2\varGamma(z)\lvert \bar{u}\rvert^2}=0 . \label{sys1_with_transformation}
\end{eqnarray}
From Eqs.~(\ref{eq1}) we also have 
\begin{equation}
i\lvert \bar{u}\rvert^2_z = u^2v^*-{u^*}^2v = \bar{u}^2 e^{i2\varGamma(z)\lvert \bar{u}\rvert^2}\bar{v}^*-\bar{u}^{*2} e^{-i2\varGamma(z)\lvert \bar{u}\rvert^2}\bar{v},
\end{equation}
such that Eqs.~(\ref{sys1_with_transformation}) takes the form
\begin{eqnarray}
i\bar{u}_z-i\varGamma(z)\bar{u}(\bar{u}^{*2} e^{-i2\varGamma(z)\lvert \bar{u}\rvert^2}\bar{v}-\bar{u}^2 e^{i2\varGamma(z)\lvert \bar{u}\rvert^2}\bar{v}^*) \nonumber \\
+\bar{u}^*\bar{v}e^{-i2\varGamma(z)\lvert \bar{u}\rvert^2}+\gamma_0\lvert \bar{u}\rvert^2 \bar{u} =0, \nonumber \\
i\bar{v}_z+q\bar{v}+\frac{\bar{u}^2}{2}e^{i2\varGamma(z)\lvert \bar{u}\rvert^2}=0. \label{eq:new_8}
\end{eqnarray}

\begin{figure}[ht!]
\begin{center}
\includegraphics[scale=0.45]{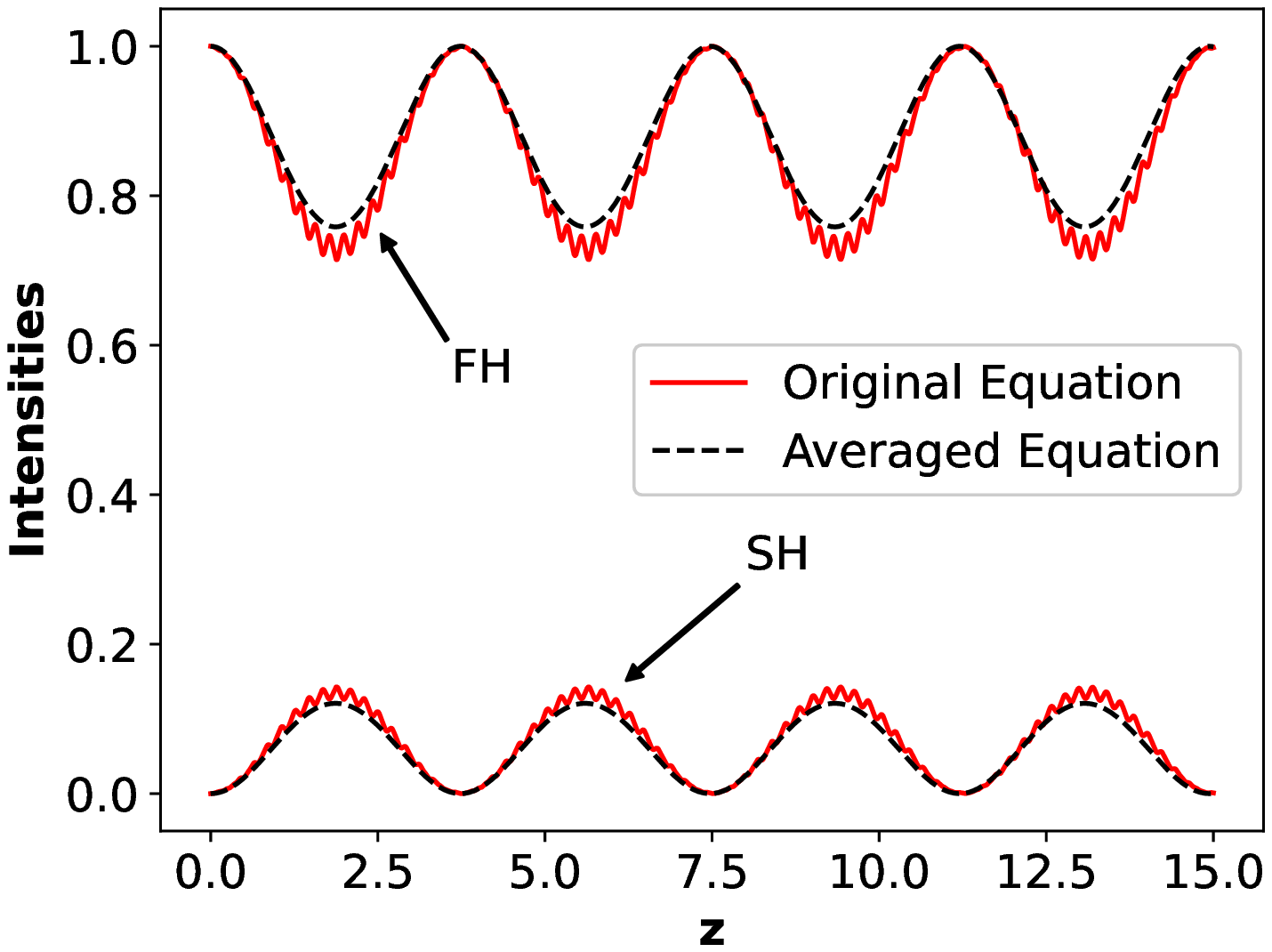}
\includegraphics[scale=0.45]{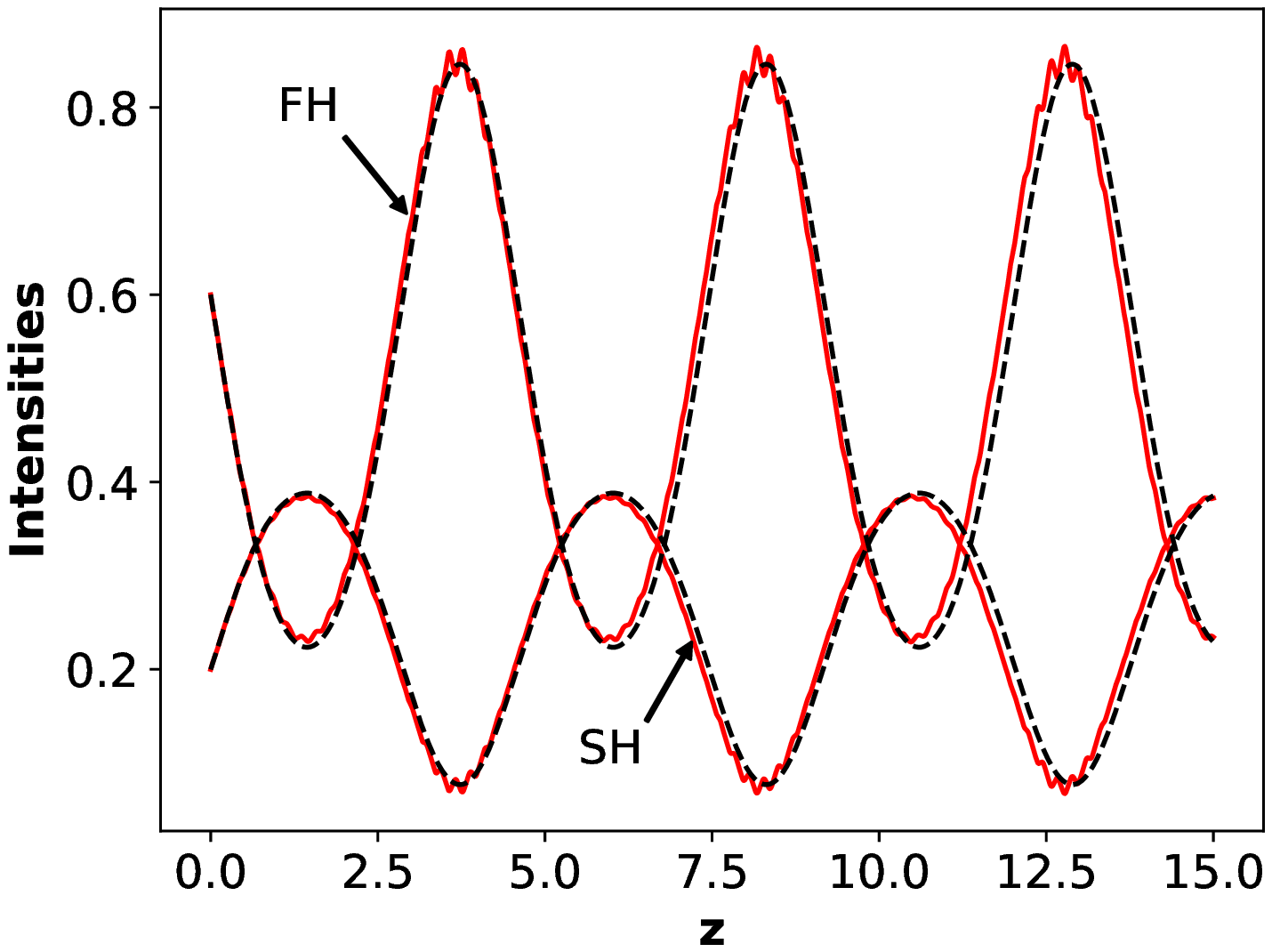}
\caption{(Color online) Comparison of solutions from the original and the averaged equations.
Numerical solutions of Eq.~(\ref{eq1}), solid (red) curves and of Eq.~(\ref{eq11}), dashed (black) curves. Parameters were here set to $\gamma_0=1, \ \gamma_1=20, \ \omega=30,$ and $q=0.1$, with initial conditions $u(0)=\bar{u}(0)=1$, $v(0)=\bar{v}(0)=0$ (upper) and $u(0)=\bar{u}(0) = \sqrt{0.6} \exp{(i\phi)}$, $v(0)=\bar{v}(0) = \sqrt{0.2}\exp{(i\psi)}$ (lower), where $\phi=0$ and $\psi=\frac{\pi}{2}$.}
\label{fig:0}
\end{center}
\end{figure}

To obtain an averaged system, we average over the period $\Lambda=2\pi/\omega$ for the rapid oscillations of Eq.~(\ref{eq3}), using the relations
\begin{eqnarray}
\langle e^{-i2\varGamma(z)\lvert \bar{u}\rvert^2}\rangle=\frac{1}{\Lambda}\int_0^{\Lambda}e^{-i2\varGamma(z)\lvert \bar{u}\rvert^2}dz=J_0 \left( \frac{2\gamma_1}{\omega}\lvert \bar{u}\rvert^2 \right), \nonumber \\
\langle \varGamma(z) e^{\pm i2\varGamma(z)\lvert \bar{u}\rvert^2}\rangle=\frac{1}{\Lambda}\int_0^{\Lambda} \varGamma(z) e^{i2\varGamma(z)\lvert \bar{u}\rvert^2}dz \nonumber \\
=\pm i\frac{\gamma_1}{\omega} J_1 \left( \frac{2\gamma_1}{\omega}\lvert \bar{u}\rvert^2 \right), \nonumber
\end{eqnarray}
where $J_i(.), i=0,1$ are  the zero and first order Bessel functions.
After inserting the averaged terms into Eqs.~(\ref{eq:new_8}), we obtain the following system of averaged coupled equations for the FH and SH
\begin{eqnarray}\label{eq11}
i\bar{u}_z - \frac{\gamma_1}{\omega} J_1 \left( \frac{2\gamma_1}{\omega}\lvert \bar{u}\rvert^2 \right) \bar{u}(\bar{u}^2 \bar{v}^{*}+\bar{u}^{*2} \bar{v}) \nonumber \\
+\bar{u}^* \bar{v} J_0 \left(\frac{2\gamma_1}{\omega}\lvert \bar{u}\rvert^2 \right) + \gamma_0\lvert \bar{u}\rvert^2 \bar{u} = 0, \nonumber \\
i\bar{v}_z+q\bar{v}+\frac{\bar{u}^2}{2}J_0 \left(\frac{2\gamma_1}{\omega}\lvert \bar{u}\rvert^2 \right)=0.
\end{eqnarray}
The Hamiltonian of the above averaged system, i.e. 
$$i\bar{u}_z=-\frac{\partial H}{\partial \bar{u}^*},  i\bar{v}_z=-\frac{\partial H}{\partial \bar{v}^*},$$
is then
\begin{equation} \label{eq:Hamiltonian}
H= \frac{\gamma_0}{2}\lvert \bar{u}\rvert^4+q\lvert \bar{v} \rvert^2+\frac{1}{2} J_0 \left( \frac{2\gamma_1}{\omega}\lvert \bar{u}\rvert^2 \right) (\bar{u}^2\bar{v}^*+\bar{u}^{*2}\bar{v}).
\end{equation}
From the above equation we conclude the important result that the Hamiltonian have the same form as 
the standard $\chi^{(2)}$-system~\cite{Buryak}, but with a renormalized effective quadratic nonlinearity parameter 
\begin{equation} \label{eq:chi_eff_parameter}
\chi_{\textnormal{eff}} = J_0 \left( \frac{2\gamma_1}{\omega}|\bar{u}|^2 \right). 
\end{equation}
Note that the renormalization depends nonlinearly on the intensity of the FH field. 
For atomic-molecular BEC systems, it means that 
the renormalized atom-molecular interaction depend nonlinearly on the atomic population.

It has previously been shown that periodic mismatch $q(z)$, and quadratic susceptibility $\chi^{(2)}(z)$, can lead to zero effective quadratic interactions, dependent on the ratio of modulation amplitudes and the frequency~\cite{Fejer,Corney}. 
Here the modulations of the cubic nonlinearity can lead to a weakening of the effective quadratic interaction, i.e.~Eq.~(\ref{eq:chi_eff_parameter}). 
A new effect seen here is that in the condition for a vanishing effective coupling ($\chi_{\textnormal{eff}}=0$), which means a zero of the Bessel function, enters the intensity of the fundamental harmonic (atomic population).

To first check the validity of the above-presented process, the original equations~(\ref{eq1}) and the corresponding averaged equations~(\ref{eq11}) have been solved numerically.
The results confirm good agreement for the averaging process, shown for two different examples of initial conditions in Fig.~\ref{fig:0}. 

\begin{figure}[ht!]
\includegraphics[scale=0.55]{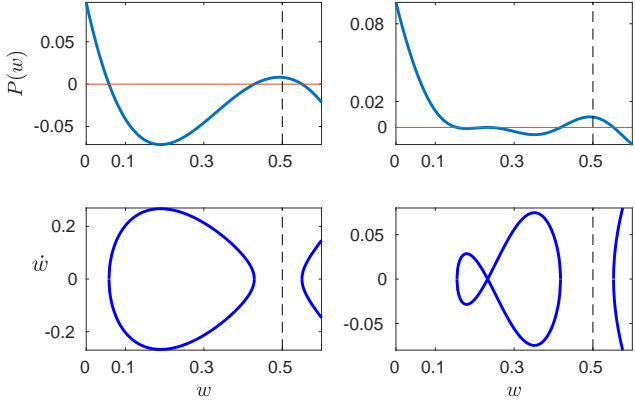}
\caption{Potential energy and phase portraits, from~Eq.~(\ref{eq:potential}). Upper and lower figures, (left) plots for the cases without modulation ($G = 0$), and (right) with modulation ($G = 4.5$). All other parameters and the Hamiltonian are the same for the four figures: $\lambda=1$, $\sigma=0.1$, and $H = -0.3107$.}
\label{fig:sadle}
\end{figure}

\begin{figure}[h]
\includegraphics[scale=0.5]{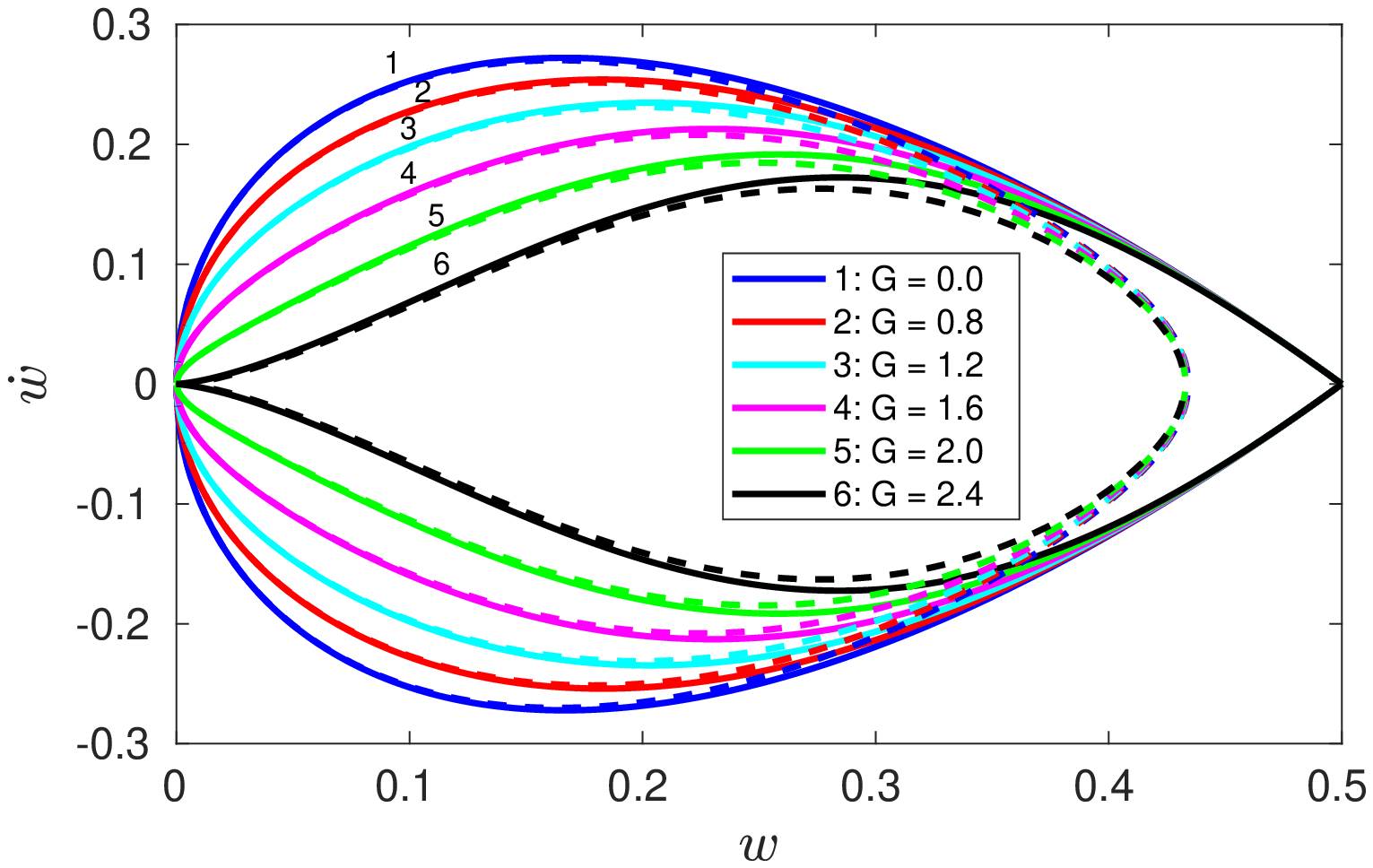}
\includegraphics[scale=0.5]{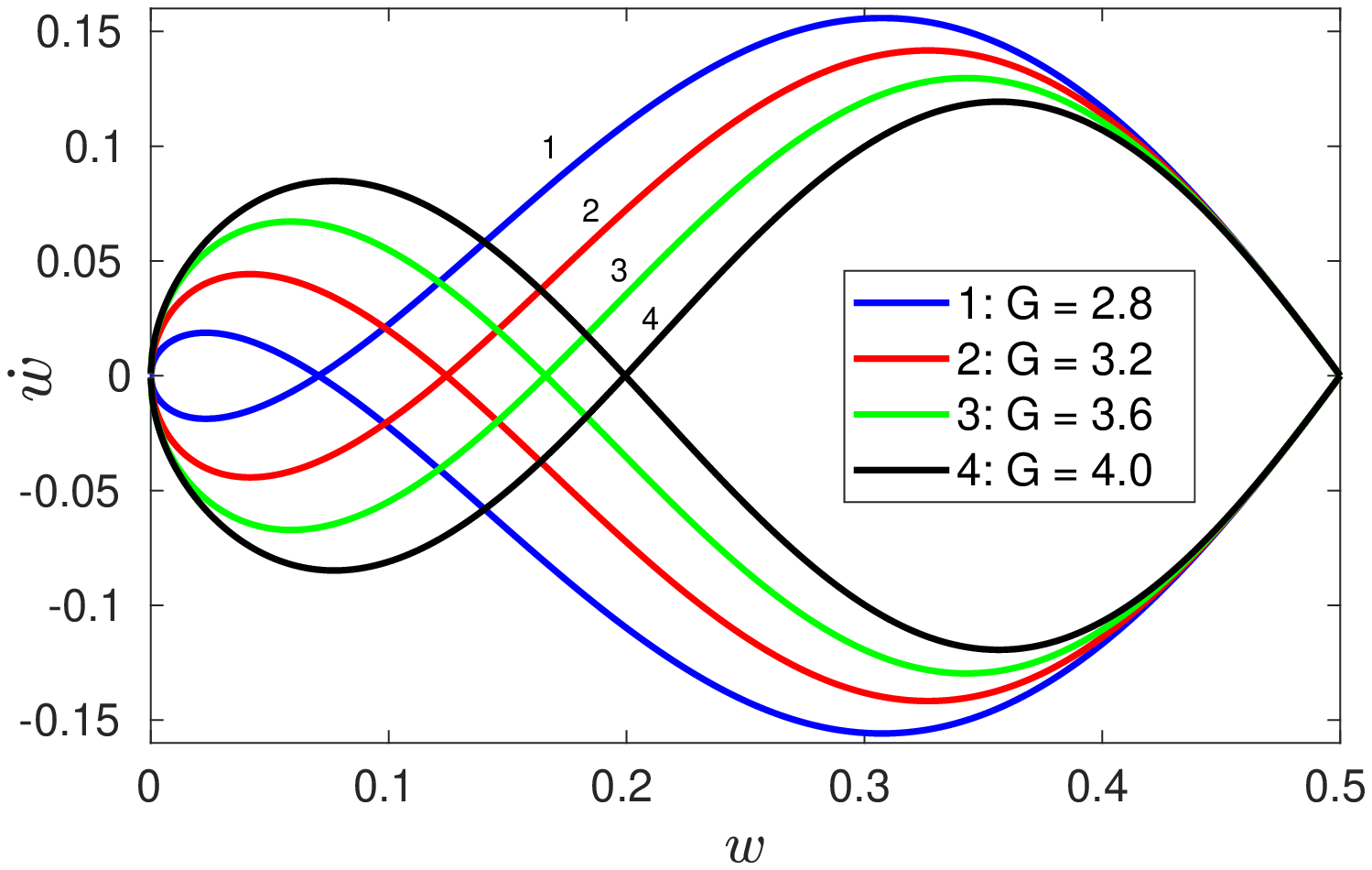}
\caption{(Color online) Phase portraits corresponding to Eq.~(\ref{gamma0=0}) for $G < G_0$ (upper), and $G \geq G_0$ (lower), where $G_0 \simeq 2.405$ is the first zero of the Bessel function $J_0$. 
In the upper portrait, the 12 curves correspond to $\sigma=0$ (solid) and $\sigma=0.10$ (dashed). 
In the lower portrait, the four curves corresponds to $\sigma=0$. 
The inset legends shows the values for $G$ for the curves 1,2,... in both portraits.
}
\label{fig:1}
\end{figure}

\section{Evolution of CW under management with rapid oscillations} \label{sec_Strong}
In this section we consider the above case of rapid modulations and obtain an equation for the intensity of the SH, i.e. $|\bar{v}|^2$, using Eqs.~(\ref{eq11}) and the averaged Hamiltonian~(\ref{eq:Hamiltonian}).
To simplify the presentation we apply the normalizations~(\ref{eq:normalizations}) for the amplitudes, the direction of propagation, and the mismatch.
After applying the normalizations above, Eq.~(\ref{eq11}) takes the following form in the new variables
\begin{eqnarray}\label{eq15}
i\rho_Z-\rho \phi_Z \nonumber \\
+L\frac{\gamma_1}{\omega}I\sqrt{I}\rho^3\mu J_1 \left( \frac{2\gamma_1}{\omega}I\rho^2 \right) \left(e^{i(2\phi-\psi)}+e^{-i(2\phi-\psi)}\right) \nonumber \\ 
+L\sqrt{I}\rho\mu J_0 \left( \frac{2\gamma_1}{\omega}I\rho^2 \right) e^{-i(2\phi-\psi)} + LI\gamma_0 \rho^3= 0, \nonumber \\
i\mu_Z-\mu \psi_Z+2\kappa \mu + L\frac{\sqrt{I}}{2}\rho^2 J_0 \left(\frac{2\gamma_1}{\omega}I\rho^2 \right) e^{i(2\phi-\psi)} = 0.
\end{eqnarray}  
Setting $G = 2\gamma_1I/\omega$,\,\, $\beta = L\sqrt{I}$, $\Upsilon = LI\gamma_0$ and $\theta=2\phi-\psi$, we have
\begin{eqnarray}
i\rho_Z-\rho \phi_Z+\beta G J_1(G\rho^2)\rho^3\mu\cos\theta \nonumber \\
+\beta J_0(G\rho^2)\rho\mu e^{-i\theta} + \Upsilon\rho^3= 0, \nonumber \\
i\mu_Z-\mu \psi_Z+2\kappa \mu + \frac{\beta}{2} J_0(G\rho^2)\rho^2 e^{i\theta} = 0. \label{eq:new_13}
\end{eqnarray}  
Splitting Eqs.~(\ref{eq:new_13}) into real and imaginary parts, setting each part separately equal to zero,
we finally have
\begin{equation}\label{eq17} 
\rho_Z=\beta J_0(G\rho^2)\rho \mu \sin\theta;  \,\,\,\,\,\,\,\, \mu_Z=-\frac{\beta}{2}J_0(G\rho^2)\rho^2 \sin\theta, 
\end{equation}
and
\begin{eqnarray}\label{eq18}
\theta_Z=2\phi_Z-\psi_Z=-2\kappa+ 2\Upsilon \rho^2 \nonumber \\
+ \left((2\beta\mu-\frac{\beta\rho^2}{2\mu})J_0(G\rho^2)+2\beta G\rho^2\mu J_1(G\rho^2)\right ) \cos\theta.
\end{eqnarray}
Eqs.~(\ref{eq17}) and (\ref{eq18}) conserve the total intensity $\rho^2 + 2\mu^2 = 1$, and also the quantity $H=2\kappa\mu^2 - 2\Upsilon(\mu^2 - \mu^4)+\beta J_0(G \rho^2)\rho^2 \mu \cos\theta $, which is related to the Hamiltonian of the system. 
It is evident from Eqs.~(\ref{eq17}) and (\ref{eq18}) that the phase evolution is determined by an interplay among periodic modulation ($G, \Upsilon$) and phase mismatch ($\beta $). 
Hence, it is convenient to introduce the parameters $\sigma=\kappa/\beta = q/(2\sqrt{I})$ and $\lambda = \Upsilon/\beta = \gamma_0\sqrt{I}$ that depends on the total intensity. 
From Eqs.~(\ref{eq17}) and~(\ref{eq18}) we can obtain the differential equation $\dot{w}=\pm \sqrt{f(w)}$, where $w = \mu^2$ is the relative intensity of the SH wave and $f(w)$ can be considered as a potential $P(w)$, like in Eq.~(\ref{2:eq8}), but with a different form 
\begin{eqnarray}\label{eq:potential}
P(w) &=& \big [H-2\sigma w+2\lambda (w-w^2)\big]^2 \nonumber \\
&-& J_0^2(G(1-2w))(1-2w)^2w=0 .
\end{eqnarray}
In Fig.~\ref{fig:sadle} the potential $P(w)$ of~(\ref{eq:potential}) and the corresponding phase portraits are depicted, in the case without modulation ($G = 0$) (left column), and with modulation ($G = 4.5$) (right column). 
We observe that the bifurcation {\it center-node} occurs when the parameter $G$ is changed. 
The simulation of the original system~(\ref{eq1}) confirms this observation.

In what follows, we focus on the relevant case without an initial SH wave (i.e., $\mu (0) = 0$), or in the atomic-molecular BEC system, no molecules initially. 
We then have $H = 0$, and obtain the following expression for $f(w)$
\begin{widetext}
\begin{equation}\label{eq19}
\dot{w} = \sqrt{f(w)} = \sqrt{J_0^2(G(1-2w))(1-2w)^2w - 4\lambda^2w^4 + 8\lambda(\lambda-\sigma)w^3 - 4(\lambda-\sigma)^2w^2}, 
\end{equation}  
\end{widetext}
where we remind the reader that an overdot denotes differentiation with respect to $\xi = \beta Z $. 
We now assume that $\lambda=0$ ($\gamma_0=0$) and then Eq.~(\ref{eq19}) takes the following form
\begin{equation}\label{gamma0=0}
\dot{w} = \sqrt{f(w)} = \sqrt{J_0^2(G(1-2w))(1-2w)^2w - 4\sigma^2 w^2}. 
\end{equation}
Eq.~(\ref{gamma0=0}) is valid when the following condition is fulfilled
\begin{equation}\label{eq20}
\lvert J_0(G(1-2w))(1-2w) \rvert \geq \lvert 2\sigma\sqrt{w}\rvert.
\end{equation}

\begin{figure}[h] 
\includegraphics[scale=0.45]{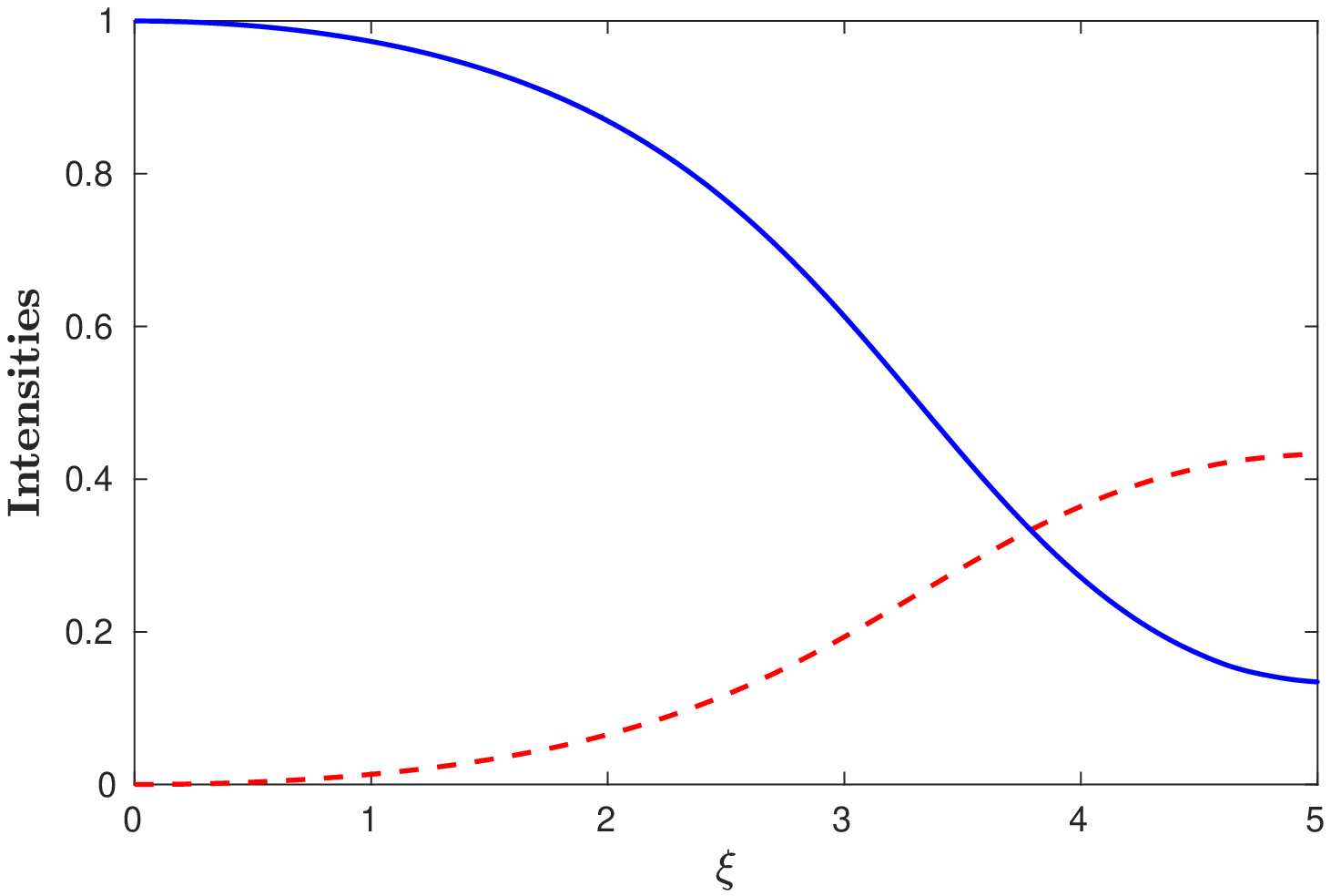}
\includegraphics[scale=0.45]{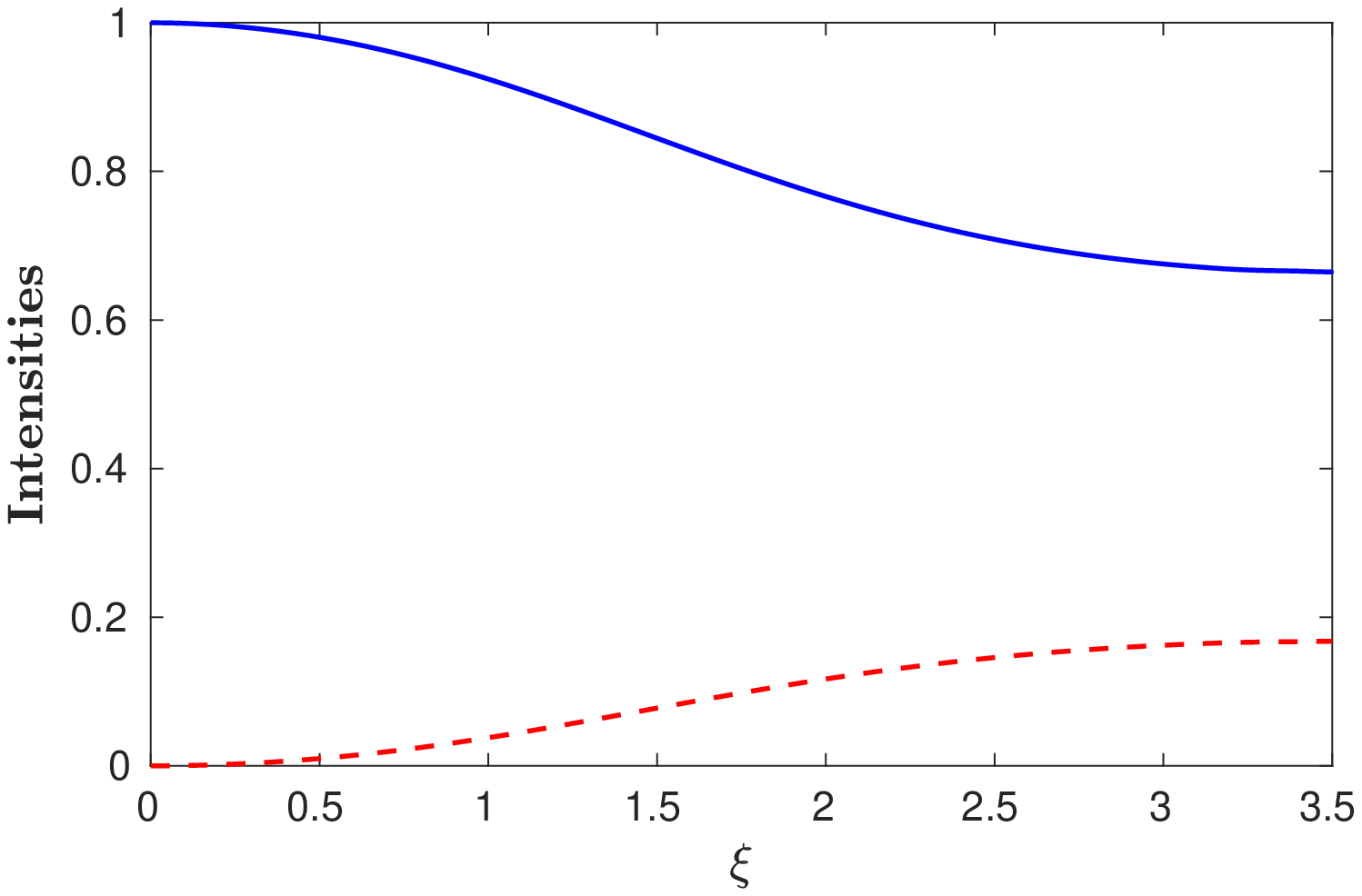}
\caption{(Color online) Numerical evolution from Eq.~(\ref{gamma0=0}) for the intensities corresponding to the case $G = 2$ (upper) and $G = 4$ (lower), with $\sigma = 0.1$ in both cases, and the initial conditions $\mu(0)=0, \: \rho(0)=1$. 
The solid (blue) curves shows the relative intensity of the FH wave ($\rho^2$), and the dashed (red) curves shows the relative intensity of the SH wave ($\mu^2$).
In the original equations these parameters corresponds to $\gamma_1/\omega = G/(2I)$ and $q=2\sigma \sqrt{I}$ where~$I=1$.}
\label{fig:2}
\end{figure}

\begin{figure}[h] 
\begin{center}
\includegraphics[scale=0.45]{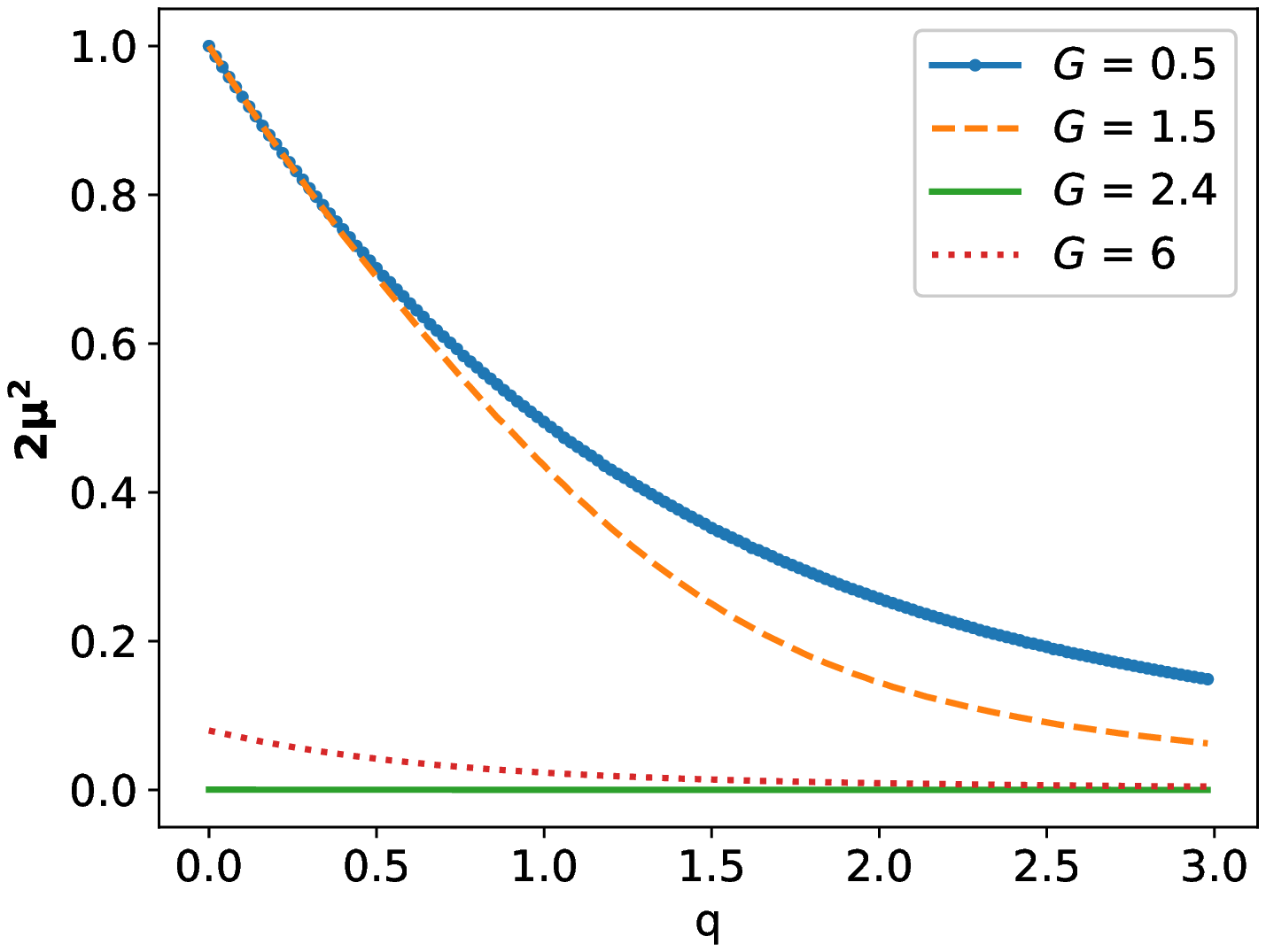}
\includegraphics[scale=0.45]{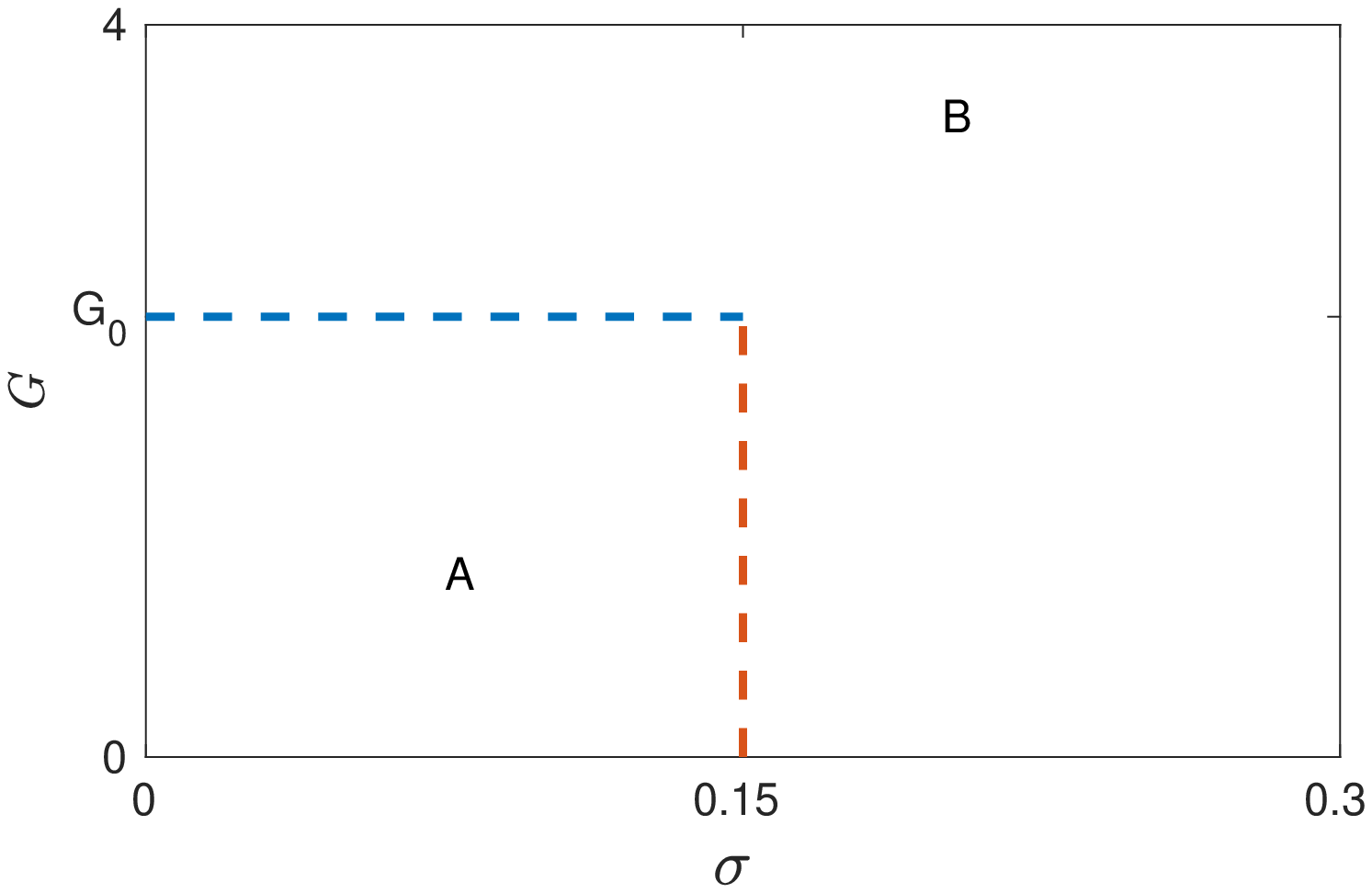}
\caption{(Color online) Ratio of the intensity of the SH wave and the total intensity (upper), as a function of $q$ ($q = 2\sigma \sqrt{I}$), for different values of $G$ ($G = 2\gamma_1 I /\omega$), see inset legend. 
Illustrative regions in the parameter plane~($\sigma, G $) obtained from the condition~(\ref{eq20}) (lower),
where $G_0 \simeq 2.405$ is the first zero of the Bessel's function $J_0$.  
More than $80 \%$ transformation of the energy defines region $A$, and less than $80 \%$ transformation defines region $B$.}
\label{fig:3}
\end{center}
\end{figure}

Fig.~\ref{fig:1} shows phase portraits obtained numerically from Eq.~(\ref{gamma0=0}), and Fig.~\ref{fig:2} shows the numerical results of integrating Eq.~(\ref{gamma0=0}), i.e. the relative intensities, $w(z)=\mu^2=\lvert \bar{v} \rvert^2/I$ and $1-2w(z)=\rho^2=\lvert \bar{u} \rvert^2/I$.
We further calculated numerically the percentage of the intensity of the SH wave, with respect to the total intensity, for different values of the parameters $\sigma$ and $G$, see Fig.~\ref{fig:3} (upper), while Fig.~\ref{fig:3} (lower) illustrates the regions of parameters corresponding to $2\mu^2 \geq 0.8$ (region $A$), and $2\mu^2 < 0.8$ (region $B$).

Let us finally comment that in the case of rapid modulations, but with weak management $2\gamma_1\lvert \bar{u}\rvert^2 / \omega \ll 1$, similar expressions for the intensities can be obtained by applying Taylor approximations to the Bessel functions $J_0$ and $J_1$ in Eqs.~(\ref{eq11}).

\section{Dynamics for slow modulation}
\label{sec_slow_modulation}

\begin{figure}
\includegraphics[scale=0.45]{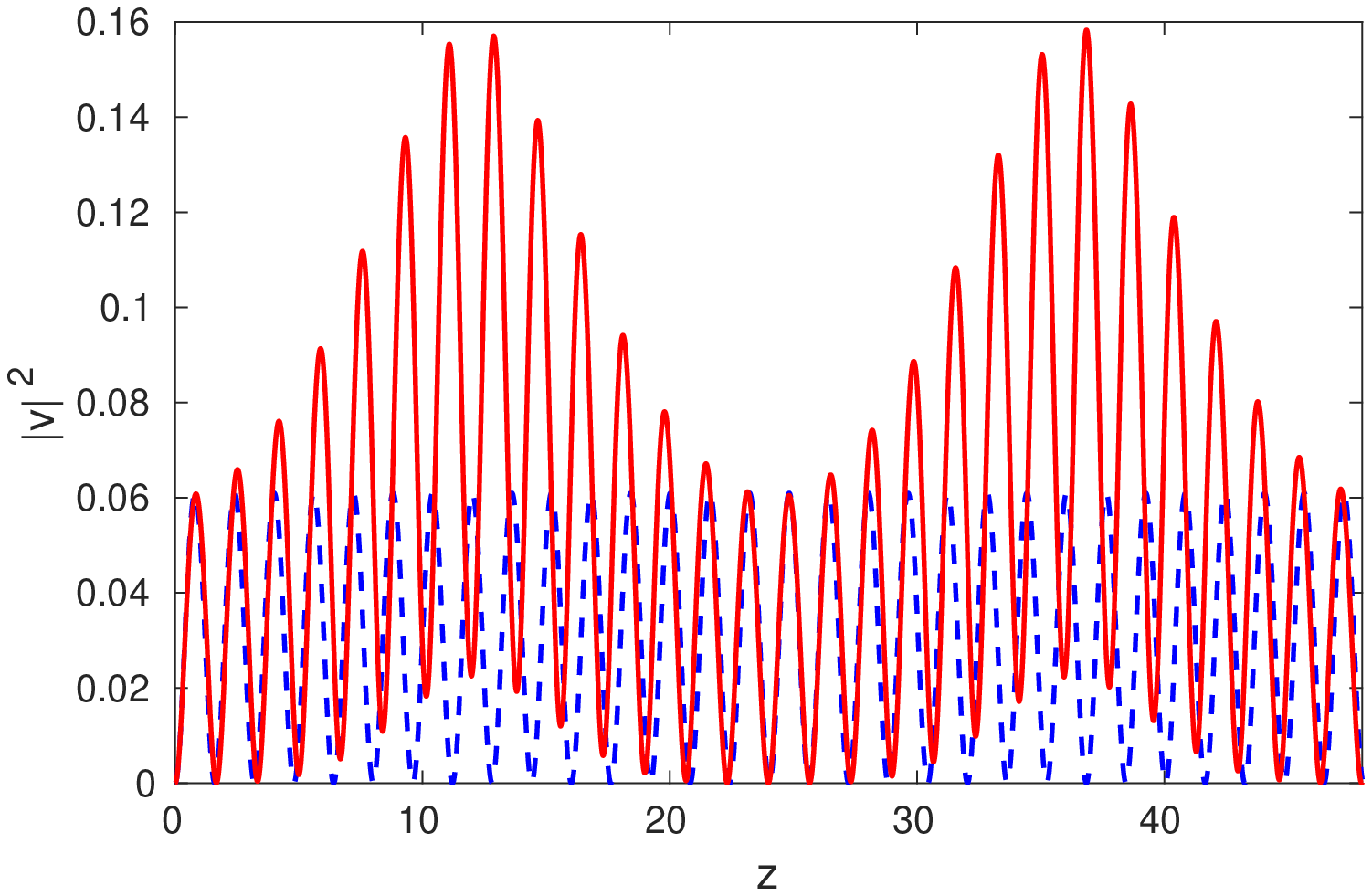}
\includegraphics[scale=0.45]{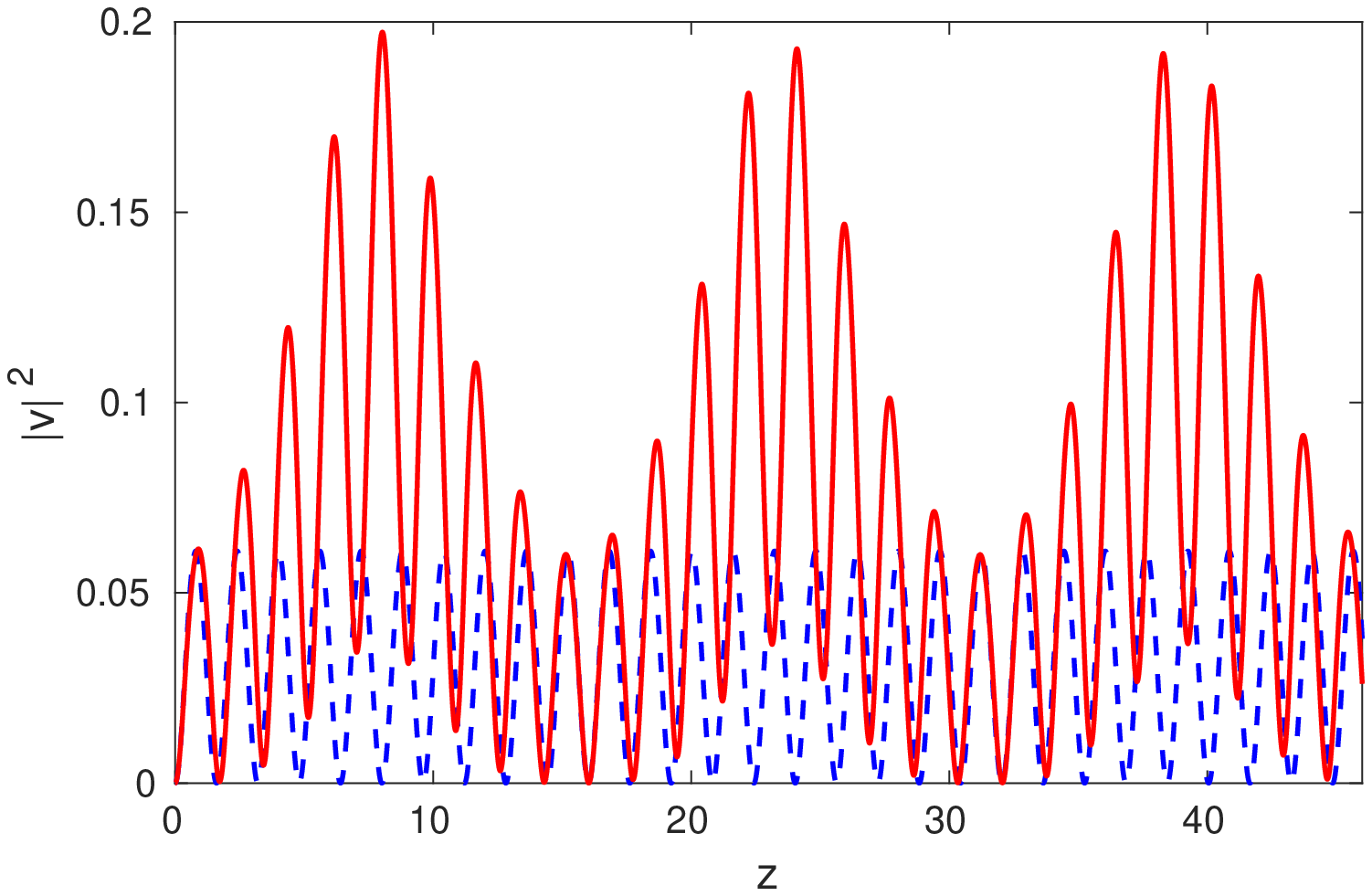}
\caption{(Color online) Numerically calculated intensity of the SH wave for resonant modulations, from Eq.~(\ref{eq1}). Dashed (blue) curves shows oscillations of the SH wave in the case where the modulation is absent ($\gamma_1=0$). Solid (red) curves shows resonance behavior when the frequency ($\omega$) of modulations for the cubic (Kerr) nonlinearity is equal to the frequency ($2r$) of oscillations for the SH wave without modulation. For the upper graph $\gamma_1 = 0.2$, and for the lower graph $\gamma_1 = 0.4$. The other parameters were $\omega=2r\approx3.95$, $\gamma_0 = 2$, $q = 0.2$, and $I = 1$ ($w_0=0.0$, $w_1=0.06090$, $w_2=0.7878$, $w_3=1.3011$).}
\label{fig:8}
\end{figure}

Let us finally study the dynamics of the system under slow resonant modulations. 
By choosing parameters so that $k \ll 1 $ and $a \ll 1$ in the solution (\ref{eqfor_w}), we can write Eq.~(\ref{eqfor_w}) in the following approximative form
\begin{equation} \label{eq:w_slow_approx}
w(\xi) \approx w_3 a \sin^2(r\xi) = \frac{w_3 a }{2}(1-\cos(2r\xi)).
\end{equation}

To check the resonant behavior in the FH to SH oscillations, we take the frequency for the modulation of the cubic (Kerr) nonlinearity to be equal to the frequency of Eq.~(\ref{eq:w_slow_approx}), i.e. $\omega= 2r$.
Results of the numerical integration are shown in Fig.~\ref{fig:8}. 
We observe resonant enhancement for the amplitude of the oscillations for the second harmonic (the molecular field) generation, which is growing with $\gamma_1$.

\begin{figure}
\includegraphics[scale=0.45]{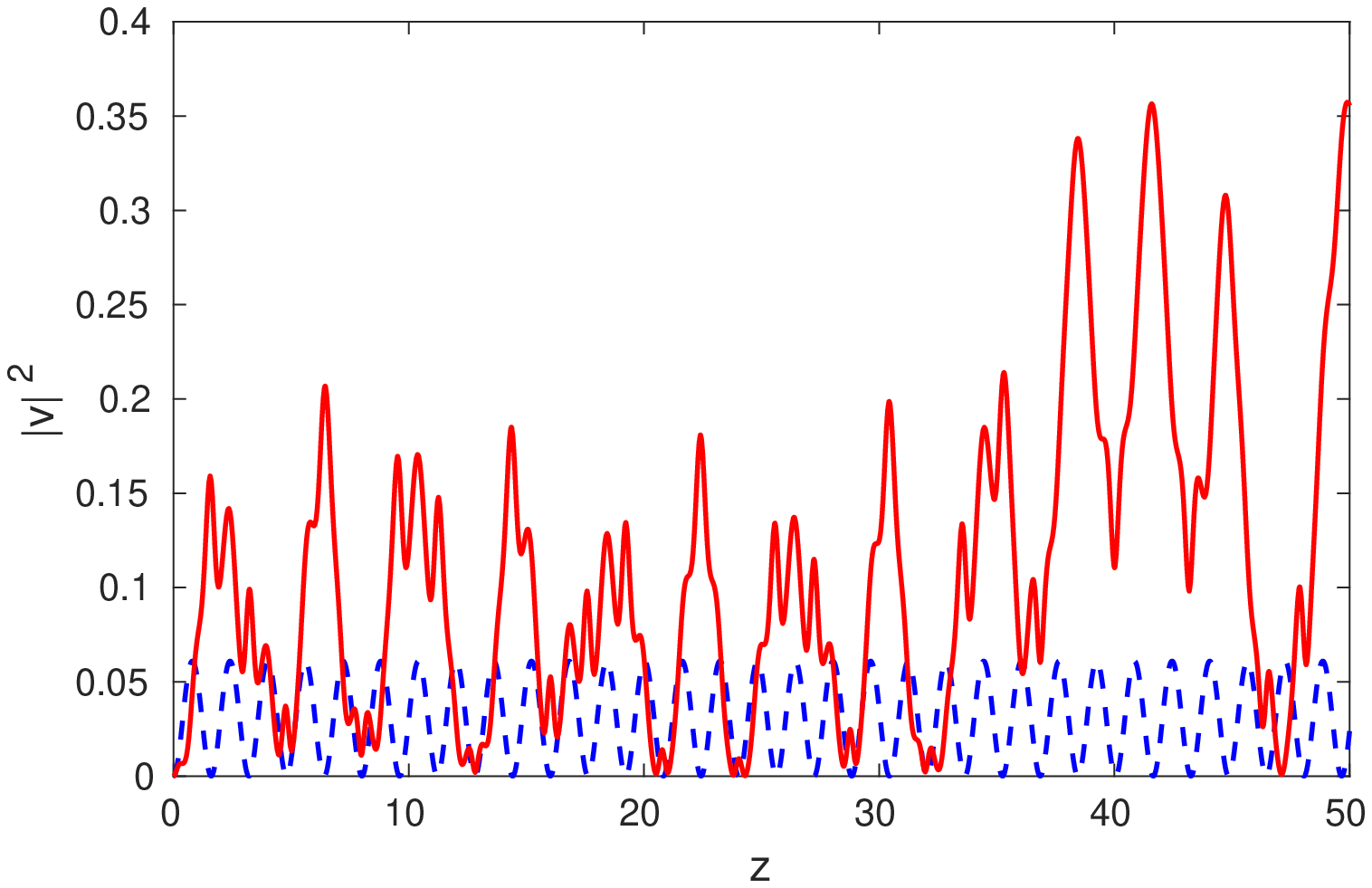}
\includegraphics[scale=0.45]{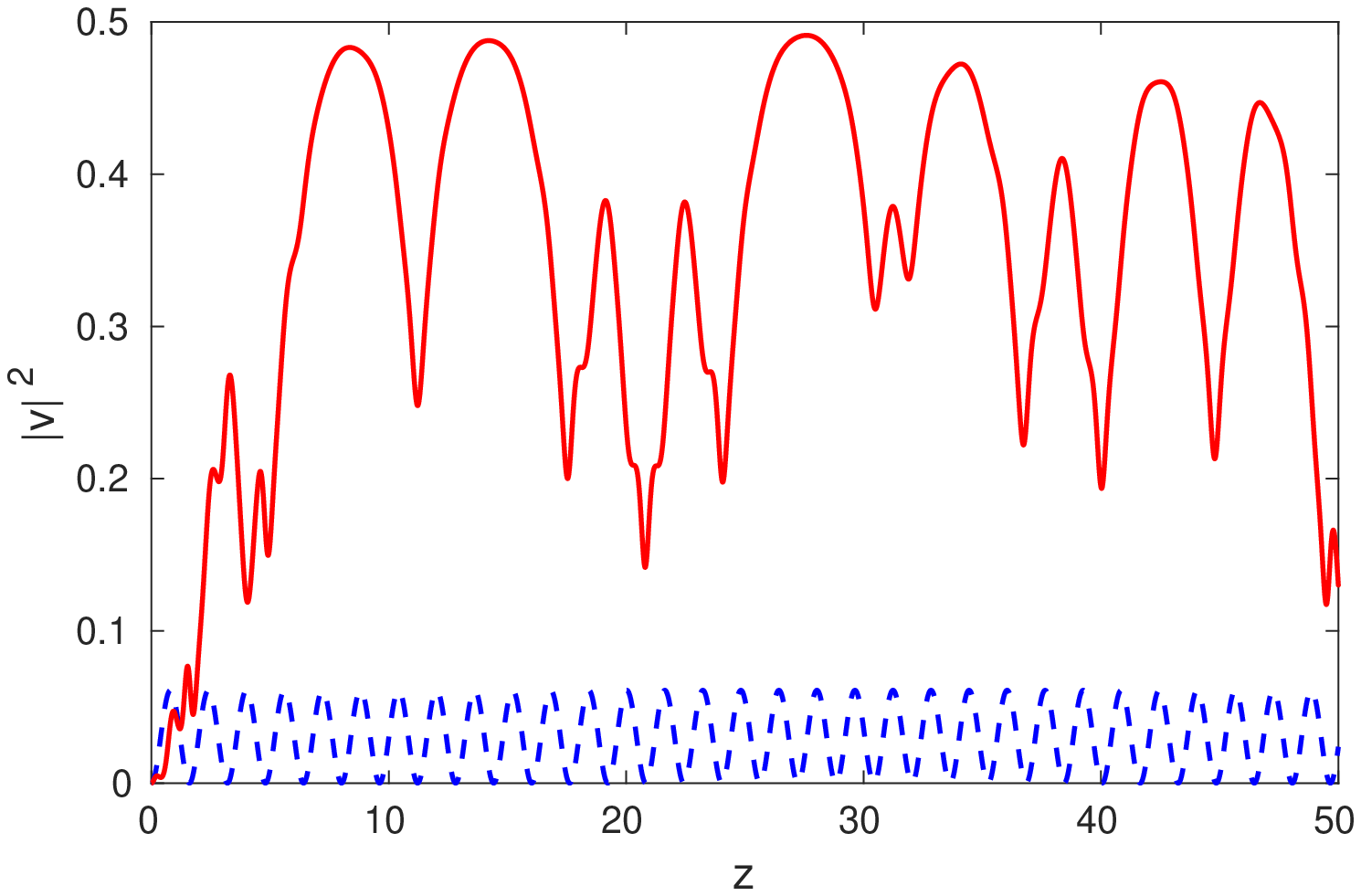}
\caption{(Color online) Numerically calculated intensity of the SH wave for strong resonant modulations, from Eq.~(\ref{eq1}). For example, for the case of $\gamma_1=5$ (upper) and $\gamma_1=6$ (lower), chaos occurs. All the other parameters are the same as in Fig.~\ref{fig:8}.}
\label{fig:9}
\end{figure}

Resonance and chaos in the second harmonic generation when the mismatch $q$ is a periodic function of $z$ has been shown in the work~\cite{Trillo}.
Chaotic oscillations originating from homoclinic crossing~\cite{Holmes,Gucken-Holmes} are also possible here. 
To investigate possible chaotic regimes of oscillations, we calculate the Melnikov function~\cite{Holmes} for the particular  case $\gamma_0=0$
(corresponding to a periodic Feshbach resonance management close to zero in the atomic scattering length)
\begin{equation} \label{eq:Melnikov_function}
M(z_0)=\int_{-\infty}^{\infty} \left( \frac{\partial H_0}{\partial w}\frac{\partial H_{I}}{\partial\theta}-\frac{\partial H_{I}}{\partial w}\frac{\partial H_0}{\partial\theta} \right) dz,
\end{equation}
where
$$
H_0 = 2\sigma w+(1-2w)\sqrt{w}\cos(\theta),\,\ H_I = -2\gamma_1 \cos(\omega z)(w-w^2).
$$
The integral is calculated on the separatrix solution of the unperturbed problem ($\gamma_1=q=0$) with $\theta=\pm\frac{\pi}{2}$, $\sin(\theta)=\pm 1$, i.e. 
\begin{equation}
\rho(z) = \mbox{sech} \left(\frac{z+z_0}{\sqrt{2}} \right), \,\,\, \mu(z)=\frac{1}{\sqrt{2}}\tanh \left(\frac{z+z_0}{\sqrt{2}}\right).
\end{equation}
The result for the Melnikov function~(\ref{eq:Melnikov_function}) is
\begin{equation} 
M(z_0) = -\frac{\pi \gamma_1}{3}\frac{\omega^2(2+\omega^2)}{\sinh(\frac{\pi\omega}{\sqrt{2}})}\sin(\omega z_0).
\end{equation}
Since the Melnikov function above has an infinite number of zeros, chaos in the harmonic generation is expected to occur~\cite{Holmes}.
This is also found numerically; see Fig.~\ref{fig:9}.

\section{Conclusion} \label{sec_Conclusion}
In this paper we have investigated the process of wave propagation in quadratic nonlinear media with an additional Kerr nonlinearity. 
Such systems can be realized in nonlinear optics and with atomic-molecular BECs. 
The Kerr nonlinearity is assumed to be periodically modulated in the direction of evolution. 
We studied both the cases of rapid and slow variations on the SH generation (the molecular field in the case of atomic-molecular BECs), and the transformation into the FH (the atomic field). 
We have derived the averaged over rapid modulations for the $\chi^{(2)}$ system with a competing Kerr nonlinearity. 
The obtained  Hamiltonian for the averaged system shows that the result for rapid modulations of the Kerr nonlinearity leads to a nonlinear renormalization of the $\chi^{(2)}$ nonlinearity coefficient.
In result, we have obtained the parameters of modulations for which the SH generation (the association of atoms into a molecular condensate) can be suppressed dynamically. 
For the case of slow modulations, we find an enhancement of the SH generation (the molecular field) for the resonant value of the frequency for the modulations of the nonlinearity, which for strong amplitudes are chaotic.
A sequential application of enhancing and suppressing modulations may be used in producing molecules from atoms.

\end{document}